\documentclass[prl,aps,twocolumn,floatfix,superscriptaddress,preprintnumbers]{revtex4-1}

\pdfoutput=1
\usepackage{mathrsfs}
\usepackage{amsfonts}
\usepackage{amsmath,amssymb,bm,bbm}
\usepackage[colorlinks=true,linkcolor=blue,citecolor=blue,urlcolor=blue]{hyperref}
\usepackage{epsfig}
\usepackage{graphicx}
\usepackage{slashed}
\usepackage{booktabs}
\usepackage{tabu}
\usepackage[dvipsnames]{xcolor}
\usepackage[normalem]{ulem}
\usepackage{tikz}
\usepackage{soul}
\usetikzlibrary{shapes,arrows}
\usetikzlibrary{positioning}
\usepackage{ulem}

\usepackage{rotating}

\newcommand{\fref}[1]{Fig.~\ref{f.#1}}

\newcommand{\eref}[1]{Eq.~(\ref{e.#1})}

\newcommand{\diff}[1]{\mathrm{d}#1}

\newcommand{\df}{\mathrm{d}}
 
\newcommand\muQ{\mu_Q}

\newcommand\bstarsc{b_*}
\newcommand\mubstar{\mu_{\bstarsc}}

\begin{document}

\preprint{JLAB-THY-23-3749, ADP-23-03/T1212}
\title{Tomography of pions and protons via transverse momentum dependent distributions}

\author{P. C. Barry}
\affiliation{Jefferson Lab,
	     Newport News, Virginia 23606, USA}
\affiliation{Physics Division, Argonne National Laboratory, Lemont,             Illinois 60439, USA}
\author{L. Gamberg}
\affiliation{Division of Science, Penn State Berks,
        Reading, Pennslyvania, 19610, USA}
\author{W. Melnitchouk}
\affiliation{Jefferson Lab,
	     Newport News, Virginia 23606, USA}
\affiliation{CSSM and CDMPP, Department of Physics, University of Adelaide,
	     Adelaide 5000, Australia}
\author{E. Moffat}
\affiliation{Division of Science, Penn State Berks,
        Reading, Pennslyvania, 19610, USA}
\affiliation{Physics Division, Argonne National Laboratory, Lemont,             Illinois 60439, USA}
\author{D. Pitonyak}
\affiliation{Department of Physics, Lebanon Valley College, Annville, Pennsylvania 17003, USA}  
\author{A. Prokudin}
\affiliation{Jefferson Lab,
	     Newport News, Virginia 23606, USA}
\affiliation{Division of Science, Penn State Berks,
        Reading, Pennslyvania, 19610, USA}
\author{N. Sato}
\affiliation{Jefferson Lab,
	     Newport News, Virginia 23606, USA}

\collaboration{{\bf 
        Jefferson Lab Angular Momentum (JAM) Collaboration}}

\begin{abstract}
We perform the first simultaneous extraction of parton collinear and transverse degrees of freedom from low-energy fixed-target Drell-Yan data in order to compare the transverse momentum dependent (TMD) parton distribution functions (PDFs) of the pion and proton. 
We demonstrate that the transverse separation of the quark field encoded in TMDs of the pion is more than $4 \sigma$ smaller than that of the proton.
Additionally, we find the transverse separation of the quark field decreases as its longitudinal momentum fraction decreases.
In studying the nuclear modification of TMDs, we find clear evidence for a transverse EMC effect.
We comment on possible explanations for these intriguing behaviors, which call for a deeper examination of tomography in a variety of strongly interacting quark-gluon systems.

\end{abstract}

\date{\today}
\maketitle

{\it Introduction.}---\
Hadrons compose nearly all the visible matter in the universe, yet much is still unknown about them.  
Revealing their internal structure from experimental data requires the use of sophisticated theoretical frameworks based on quantum chromodynamics (QCD), the theory of the strong force of quarks and gluons (partons), that describe hadrons as emergent phenomena. 
While decades of high-energy experiments have provided data allowing for the high resolution of the longitudinal structure of protons~\cite{Dulat:2015mca, NNPDF:2017mvq, Alekhin:2017kpj, Hou:2019efy, Bailey:2020ooq, Moffat:2021dji}, and to a lesser extent also of pions~\cite{Owens:1984zj, Aurenche:1989sx, Sutton:1991ay, Gluck:1991ey, Gluck:1999xe, Wijesooriya:2005ir, Aicher:2010cb, Barry:2018ort, Novikov:2020snp, Cao:2021aci, Barry:2021osv, Barry:2022aix}, the information on the transverse structure of hadrons is comparatively less well known.
In particular, achieving a 3-dimensional mapping of internal hadron structure requires sensitivity to both collinear and transverse parton degrees of freedom, which can be encoded in transverse momentum dependent distributions (TMDs)~\cite{Kotzinian:1994dv, Tangerman:1994eh, Mulders:1995dh, Boer:1997nt, Collins:2011zzd} and generalized parton distributions (GPDs)~\cite{Diehl:2003ny, Belitsky:2005qn}. 
Both are primary focuses at existing and future facilities, such as Jefferson Lab~\cite{Dudek:2012vr} and the Electron-Ion Collider~\cite{AbdulKhalek:2021gbh}.
Here we focus on TMDs and novel properties of the transverse separation of quark fields as a function of their longitudinal momenta for the proton and pion, giving deeper insights into color confined systems that emerge from QCD.

TMD parton distribution functions (PDFs) depend on both the longitudinal momentum fraction $x$ and the intrinsic transverse momentum  $k_T \equiv |{\bm k_T}|$ of partons inside the hadron.
The unpolarized TMD PDF is the $k_T$-space Fourier transform of the following light-front correlator of hadron ${\cal N}$ (with momentum $P$)~\cite{Mulders:1995dh, Collins:2011zzd}, 
\begin{align}
    	&\tilde f_{q/{\cal N}} (x,b_T) \label{e.correlator} \\ &= \int\frac{ \diff b^-}{4\pi}
	\,e^{-i x P^+ b^-} {\rm Tr}\big[ \langle\, {\cal N}\, |\, \bar \psi_q(b) 
	\gamma^+  \mathcal{W}(b,0) \psi_q(0)\, |\, {\cal N}\, \rangle \big] ,\nonumber
\end{align}
where $b\equiv (b^-,0^+,{\bm b}_T)$, with ${\bm b_T}$ the transverse shift of the quark field $\psi_q$, and $b_T \equiv |{\bm b}_T|$.
The Wilson line $\mathcal{W}(b,0)$ ensures SU(3) color gauge invariance and is understood here to be the staple-shaped gauge link for the Drell Yan process~\cite{Collins:2002kn}.
The correlator in \eref{correlator} requires a modification to account for the ultraviolet and rapidity divergences, and acquires corresponding regulators $\tilde{f}_{q/{\cal N}}(x,b_T) \to \tilde{f}_{q/{\cal N}}(x,b_T;\mu,\zeta)$~\cite{Aybat:2011zv, Collins:2011zzd}.

While $\tilde f_{q/{\cal N}}$ is technically the object to be inferred from data, its small-$b_T$ behavior can be written in terms of collinear PDFs~\cite{Collins:1984kg, Collins:2011zzd, Collins:2014jpa}.
Most phenomenological extractions to date~\cite{Bacchetta:2017gcc, Bacchetta:2019sam, Bertone:2019nxa, Scimemi:2019cmh, Bacchetta:2022awv, Wang:2017zym, Vladimirov:2019bfa, Cerutti:2022lmb} have made use of this connection by fixing the collinear PDFs and focusing on the analysis of the nonperturbative large-$b_T$ region.
However, such extractions are subject to the choices of the input collinear PDFs, as discussed in Ref.~\cite{Bury:2022czx}.

In this Letter, we go beyond previous studies by performing the first simultaneous extraction of proton and pion TMD PDFs, along with pion collinear PDFs, through an analysis of fixed-target Drell-Yan (DY) and leading neutron (LN) data within the JAM QCD analysis framework~\cite{Jimenez-Delgado:2013sma, Jimenez-Delgado:2013boa, Jimenez-Delgado:2014xza, Sato:2016tuz, Sato:2016wqj, Ethier:2017zbq, Lin:2017stx, Barry:2018ort, Sato:2019yez, Cammarota:2020qcw, Bringewatt:2020ixn, Moffat:2021dji, Adamiak:2021ppq, Cao:2021aci, Cocuzza:2021rfn, Zhou:2021llj, Barry:2021osv, Cocuzza:2021cbi, Zhou:2022wzm, Boglione:2022gpv, Cocuzza:2022jye, Barry:2022aix, Gamberg:2022kdb}.
We find an intriguing behavior of the average transverse separation of the quark fields encoded in these TMD PDFs, with $\sim (4 - 5.2)\, \sigma$ smaller values for the pion than for the proton as a function of~$x$. 
For both systems the average transverse separation of the quark field decreases as $x$ decreases.
We also observe a transverse EMC effect by studying the nuclear modification of the TMDs, with the average transverse separation of a quark field in a bound proton up to 12\% smaller than that in a free proton.

{\it Analysis framework.}---
The focus of our analysis is the DY process in hadron-hadron or hadron-nucleus reactions with center of mass energy~$\sqrt{s}$.
Specifically, we study the region of small transverse momentum $q_T$ of the produced lepton pair relative to its invariant mass~$Q$.
In this regime, the measured cross section can be described through a rigorous factorization framework in terms of TMD PDFs in $b_T$-space~\cite{Collins:1981uk,Collins:1981va,Collins:1984kg,Ji:2004xq,Becher:2010tm,Collins:2011zzd,Echevarria:2011epo}.
The DY cross section for hadron ($\mathcal{N}=\pi, p$)--nucleus ($A$) collisions differential in $Q^2$, rapidity $y$ of the lepton pair, and $q_T$ is given~by  
\begin{align}
 \frac{\df^3\sigma}{\df Q^2 \df y \df q_T^2} 
 & = \sum_{q}  \mathcal{H}_{q\bar q}^{{\mbox{\tiny \rm DY}}}(Q,\muQ) 
  \int\! \frac{\df^2{\bm b}_T}{(2\pi)^2} \, e^{i {\bm b}_T \cdot {\bm q}_T} \nonumber \\
  &\hspace*{-1cm}\times \tilde f_{q/\mathcal{N}}(x_\mathcal{N}, b_T; \muQ, Q^2) \,
   \tilde f_{\bar q/A}(x_A, b_T; \muQ, Q^2)
\label{e.DYpTxsec}
\,.
\end{align}
Here, the hard factor $\mathcal{H}^{{\mbox{\tiny \rm DY}}}_{q\bar q}$ represents the process-dependent perturbatively calculable hard scattering that is factorized from the process-independent TMDs $\tilde{f}_{q(\bar{q})/{\mathcal N} (A)}$.
The longitudinal momentum fractions of the TMDs are kinematically constrained to be $x_{\mathcal{N}(A)} = \sqrt{\tau}e^{+(-)y}$, where $\tau = Q^2/s$.
Additionally, to optimize the perturbative calculation, the scale dependence is set as $\mu_Q=Q$ and the rapidity scale $\zeta=Q^2$~\cite{Collins:2011zzd}.

We use the standard $b_*=b_T/\sqrt{1+b_T^2/b_{\rm max}^2}$ prescription to model the large-$b_T$ behavior of the TMDs, and following Ref.~\cite{Collins:2017oxh} we take
\begin{align}
\label{e.tmd}
&\tilde f_{q/{\cal N}(A)}(x, b_T; \muQ, Q^2) =   
(C\otimes f)_{q/{\cal N}(A)}(x;b_*)  \\
 &\!\times\!  \exp\!\left\{\!\! 
  -g_{{q/{\cal N}(A)}}(x,b_T)\!-\!g_{K}(b_T) \ln\! \frac{Q}{Q_0}\! - \!S(b_*,Q,\mu_Q)
  \!\right\}. \nonumber
\end{align}
Here, the first line is the operator product expansion (OPE)~\cite{Collins:1984kg, Collins:2011zzd, Collins:2014jpa}, which describes the small-$b_T$ behavior of the TMDs in terms of the collinear PDFs $f_{q/\mathcal{N}(A)}$ convoluted with perturbative Wilson coefficients~$C$.
The first two terms in the second line are  nonperturbative functions to be extracted from experiment: $g_{q/{\cal N}(A)}$~\cite{Collins:2011zzd} that describes the deviation from the OPE at large $b_T$, and the nonperturbative part $g_K$ of the Collins-Soper (CS) kernel~\cite{Collins:2011zzd}. 
The factor $S$ contains the perturbative effects of soft gluon radiation, which can be written as
\begin{align}
  \label{eq:S}
  & S(b_*,Q,\mu_Q)
  = 
  - \widetilde{K}(b_*;\mubstar) \ln \frac{ Q }{ \mubstar } \\
  & + \int_{\mubstar}^{\mu_Q}  \frac{ d{\mu'} }{ \mu' }
           \biggl[ -  \gamma_f(\alpha_s(\mu'); 1) 
                   + \ln\frac{Q}{\mu'} \gamma_K(\alpha_s(\mu'))
           \biggr] \nonumber     \; , 
\end{align}
with $\mubstar = 2e^{-\gamma_{\rm E}}/b_*$, $\gamma_K$ the cusp anomalous dimension~\cite{Polyakov:1980ca, Korchemsky:1985xj, Korchemsky:1987wg, Collins:1989bt, Moch:2004pa, Henn:2019swt}, $\gamma_f$ the anomalous dimension of the TMD operator~\cite{Collins:2011zzd, Moch:2005id, Collins:2017oxh}, and $\widetilde K$ the perturbative part of the CS kernel~\cite{Collins:1981uk, Collins:1981uw, Collins:2011zzd, Collins:2017oxh}.
To remain consistent with the collinear observables used in this analysis, we take the hard factor $\mathcal{H}^{{\mbox{\tiny \rm DY}}}_{q\bar q}$ in \eref{DYpTxsec} and Wilson coefficients $C$ in \eref{tmd} at $\mathcal{O}(\alpha_s)$~\cite{Collins:2011zzd, Collins:2017oxh}.
Therefore, in logarithmic counting as described in Refs.~\cite{Bacchetta:2019sam, Boussarie:2023izj}, we implement next-to-next-to-leading-logarithmic (N$^2$LL) accuracy for TMD PDFs by using $\gamma_K$ at $\mathcal{O}(\alpha_s^3)$ and $\gamma_f$ and $\widetilde K$ at $\mathcal{O}(\alpha_s^2)$ precision.
The number of active flavors is determined by the hard scale $Q$. 
We use the starting scale $Q_0 = 1.27~{\rm GeV}$ and $b_{\rm max} = 2 e^{-\gamma_E}/Q_0 \approx 0.88~{\rm GeV}^{-1}$.

Since we analyze $\pi A$ and $p A$ DY data, we cannot extract TMDs for $p$, $\pi$ and $A$ systems simultaneously from two independent processes.
We therefore relate the TMD PDFs for the nucleus to bound proton and neutron TMD PDFs by the relations
$\tilde f_{q/{A}} \equiv (Z/A) \tilde f_{q/{p/A}} + (1-Z/A) \tilde f_{q/{n/A}}$, 
for a nucleus with mass number $A$ and atomic number $Z$.
In modeling the nuclear dependence, in the large-$b_T$ region we introduce an $A$ dependence in the quantity 
$g_{q/{{\cal N}/A}} = g_{q/{\cal{N}}}(1+a_\mathcal{N} (A^{1/3}-1))$,
where $a_\mathcal{N}$ is a fit parameter~\cite{Alrashed:2022jlx}.
The quantity $g_K$ is universal and does not need to be modified.
In the small-$b_T$ region controlled by nuclear collinear PDFs, we describe the quarks in the bound nucleons inside the nucleus following previous collinear nuclear PDF analyses~\cite{Eskola:2016oht, Kovarik:2015cma},
    $f_{u/p/A}$ = $[Z/(2Z-A)] f_{u/A}$ $+$ $[(Z-A)/(2Z-A)] f_{d/A}$, {\it etc.},
with $f_{q/A}$ taken from the EPPS16 analysis~\cite{Eskola:2016oht}.
To be consistent with EPPS16, we utilize the CT14 proton NLO PDFs~\cite{Dulat:2015mca} in the $pA$ reactions, and consequently use Wilson coefficients in the OPE at $\mathcal{O}(\alpha_s)$.
In principle, the analysis can depend on the choice of collinear proton PDFs, but in practice, we see minimal difference in our results (see below).
In the future we will include collider data and extract simultaneously PDFs and TMDs of the proton along with the pion, eliminating any such dependence.

In the collinear sector, our treatment of the $q_T$-integrated DY cross section included NLO accuracy in the hard coefficients~\cite{Barry:2018ort, Cao:2021aci}. 
For the LN reactions, we utilize the combined chiral effective theory and collinear factorization to describe the Sullivan process, as described in Refs.~\cite{Sullivan:1971kd, McKenney:2015xis}. 

We employ the Bayesian Monte Carlo (MC) methodology of the JAM Collaboration~\cite{Jimenez-Delgado:2013sma, Jimenez-Delgado:2013boa, Jimenez-Delgado:2014xza, Sato:2016tuz, Sato:2016wqj, Ethier:2017zbq, Lin:2017stx, Barry:2018ort, Sato:2019yez, Cammarota:2020qcw, Bringewatt:2020ixn, Moffat:2021dji, Adamiak:2021ppq, Cao:2021aci, Cocuzza:2021rfn, Zhou:2021llj, Barry:2021osv, Cocuzza:2021cbi, Zhou:2022wzm, Boglione:2022gpv, Cocuzza:2022jye, Barry:2022aix, Gamberg:2022kdb}.
To explore the model dependence of our extractions, we implement a variety of intrinsic nonperturbative functions:~Gaussian~\cite{Davies:1984sp, Nadolsky:1999kb, Landry:2002ix, Anselmino:2013lza}, exponential, an interpolation of Gaussian-to-exponential~\cite{Bertone:2019nxa, Vladimirov:2019bfa}, Bessel-like~\cite{Aidala:2014hva, Sun:2014dqm, Radyushkin:2014vla, Radyushkin:2016hsy, Boglione:2022nzq}, and a sum of Gaussians (MAP parametrization)~\cite{Bacchetta:2022awv}.
In addition, we explore parametrizing $g_K$ using Gaussian~\cite{Nadolsky:1999kb, Anselmino:2013lza, Bacchetta:2022awv}, exponential~\cite{Bertone:2019nxa, Vladimirov:2019bfa}, and logarithmic~\cite{Kang:2014zza, Aidala:2014hva, Kang:2015msa} forms at large $b_T$. 
A detailed comparison of all these parametrizations will be in a forthcoming paper~\cite{FUTUREPAPER}.

{\it Phenomenology.}---\
As noted above, we include in this analysis both $q_T$-dependent and collinear data, and are consequently able to, for the first time, simultaneously extract the pion's TMD and collinear PDFs.
Table~\ref{t.MCfits} summarizes all of the datasets included in our analysis.

We use data from the E288  experiment~\cite{Ito:1980ev} taken with 200, 300, and 400~GeV proton beams on a platinum (Pt) target, the E605  experiment~\cite{Moreno:1990sf} taken on a copper (Cu) target, and the E772 experiment~\cite{E772:1994cpf} using a deuterium target expressed as
$E\df^3\sigma/\df^3 {\bm q} = (1/\pi) \df^2 \sigma/\df y\, \df q_T^2$.
The E288 and E605 experiments took measurements at fixed rapidity and fixed $x_F = 2\sqrt{\tau}\sinh(y)$, respectively, while the E772 experiment measured over $0.1\leq x_F \leq 0.3$.

This analysis also includes the ratios
$R_{A,B}= (\df \sigma / \df q_T)|_{pA} / (\df \sigma / \df q_T)|_{pB}$ 
of DY cross section per nucleon from an 800~GeV proton beam incident on beryllium (Be), iron (Fe), and tungsten (W) targets from the E866 experiment~\cite{NuSea:1999egr}. 
These datasets are particularly sensitive to  nuclear TMDs~\cite{Alrashed:2022jlx}.
We integrate over the $Q$ range for each bin and the measured $0<x_F<0.8$ range.

To constrain the pion TMDs, we include $q_T$-dependent DY data  from the E615~\cite{Conway:1989fs} and E537~\cite{Anassontzis:1987hk} experiments.
While these also measured $\df^2 \sigma/\df Q\, \df q_T$, the observables we consider are $\df^2 \sigma/\df x_F\, \df q_T$.
In the $Q$-dependent cross section, an integration over $0<x_F<1$ would be required, the upper limit of which is not well defined in the factorization approach.
On the other hand, the range of $Q$ integration for the $x_F$-dependent cross section is well within the region where factorization is valid.
Complementary to the $q_T$-differential data, we include the pion-induced $q_T$-integrated DY data from the E615 and NA10~\cite{NA10:1985ibr} experiments measuring $\df^2\sigma/\df \sqrt{\tau} \df x_F$, which strongly constrain the pion's valence quark PDF.
We also include the LN electroproduction data from HERA~\cite{H1:2010hym, ZEUS:2002gig} as in previous JAM analyses~\cite{Barry:2018ort, Cao:2021aci, Barry:2021osv, Barry:2022aix}.

To ensure the validity of TMD factorization, we impose a cut~\cite{Vladimirov:2019bfa} on the data to small $q_T$:~$q_{T}^{\rm max}<0.2\,  Q$, where $q_{T}^{\rm max}$ is the upper bound of the $q_T$ bin.
We restrict our analysis to data in the range $4 < Q < 9$~GeV and $Q > 11$~GeV to avoid the region of $J/\psi$ and $\Upsilon$ resonances.
Following Ref.~\cite{Vladimirov:2019bfa}, we impose a cut on the $q_T$-dependent and $q_T$-integrated DY data of $x_F<0.8$ to avoid regions where threshold resummation may be additionally needed for both observables.
For the collinear DY observables, we use $4.16<Q<7.68~{\rm GeV}$.
Cuts on the LN data were imposed as in Refs.~\cite{Barry:2018ort, Cao:2021aci, Barry:2021osv, Barry:2022aix}.

In all, we analyze 67 $q_T$-dependent pion-induced and 238 proton-nucleus DY data points, 111 $q_T$-integrated pion-induced DY data points, and 108 data points from the LN experiments, for a total of 524 data points.
In exploring the various nonperturbative parametrizations, we observed that the Gaussian $g_K$ and multi-component (sum of Gaussians) MAP-like~\cite{Bacchetta:2022awv} flavor-independent parametrizations for the intrinsic $g_{q/{\cal{N}}}$ have the best agreement across all $q_T$-dependent observables by an improved $\chi^2$ per number of points ($N$) of between 0.44 and 1.92.
We do not find any significant improvement in the description of the data with the inclusion of flavor dependence.

In the end we have a total of 25 free parameters:~3 parameters for $g_{q/\pi}$ and 11 for $g_{q/p}$ plus one parameter for nuclear dependence and one parameter for $g_K$  to model the TMDs, along with an additional 8 parameters for pion collinear PDFs, and one LN cutoff parameter.
We find largely that the sensitivity of the parameter correlations is not strong between pions and protons, neither in the TMD nor the collinear regions.
The correlations are largely self-contained within each distribution, i.e., the pion PDF parameters are correlated with each other, and similarly for the proton and pion TMD parameters individually.

\begin{table}[ht]
\caption{Datasets included in this analysis, along with the resulting $\chi^2$ per datum and $Z$-scores from the MC analysis.} 
\centering
\begin{tabular}{ |l | c | c| c c|}
\hline
{Process} &
{Experiment} &
$\sqrt{s}$ (GeV) &
$\chi^2/{N}$ &
$Z$-score
\\
\hline
\hline
\multicolumn{5}{|c|}{\bf TMD} \\
\hline
{\rm $q_T$-dep. $pA$ DY} &
E288~\cite{Ito:1980ev} & 19.4 &
1.07 &
0.34
\\
$p A \to \mu^+ \mu^- X$
&
E288~\cite{Ito:1980ev} &23.8 &
0.99 &
0.05
\\
&
E288~\cite{Ito:1980ev} &  24.7&
0.82 &
0.99
\\
&
E605~\cite{Moreno:1990sf} & 38.8 &
1.22 &
1.03
\\
&
E772~\cite{E772:1994cpf} & 38.8 &
2.54 &
5.64
\\
~~~~~~(Fe/Be)~~~ &
E866~\cite{NuSea:1999egr} & 38.8 &
1.10 &
0.36
\\
~~~~~~(W/Be)~~~ &
E866~\cite{NuSea:1999egr} & 38.8&
0.96 &
0.15
\\
\hline
{\rm $q_T$-dep. $\pi A$ DY}&
E615~\cite{Conway:1989fs} & 21.8 &
1.45 & 
1.85
\\
$\pi W \to \mu^+ \mu^- X$
&
E537~\cite{Anassontzis:1987hk} & 15.3 &
0.97 & 
0.03
\\\hline \hline
\multicolumn{5}{|c|}{\bf collinear} \\
\hline
$q_T$-{\rm integr. DY} &
E615~\cite{Conway:1989fs} & 21.8 &
0.90 &
0.48
\\
$\pi W \to \mu^+ \mu^- X$
&
NA10~\cite{NA10:1985ibr} & 19.1 &
0.59 &
1.98
\\
&
NA10~\cite{NA10:1985ibr} & 23.2 &
0.92 &
0.16
\\
\hline
{\rm leading neutron} & 
H1~\cite{H1:2010hym} & 318.7~\, &
0.36 &
4.59
\\
~~~$e p \to e n X$
&
ZEUS~\cite{ZEUS:2002gig} & 300.3~\, &
1.48 &
2.15
\\
\hline
\hline
\multicolumn{3}{|l|}{Total} &
1.12 & 
1.86
\\
\hline
\end{tabular}
\label{t.MCfits}
\end{table}

The resulting agreement with data is shown in Table~\ref{t.MCfits}, where the $\chi^2/N$ and the $Z$-scores are provided for each of the experimental datasets considered.
The $Z$-score is the inverse of the normal cumulative distribution function, $Z=\Phi^{-1}(p) \equiv \sqrt{2}~{\rm erf}^{-1}(2p-1)$, where the $p$-value is computed according to the resulting $\chi^2$ shown in Table~\ref{t.MCfits}, and it describes the significance of the $\chi^2$ relative to the expected $\chi^2$ distribution.
Our analysis shows a relatively good compatibility between data and theory at the level of the $Z$-score (1.86), with a total $\chi^2/N=1.12$.
The worst agreement with the datasets was to the E772 data, which provided a $Z$-score of above~5.
Other analyses~\cite{Bertone:2019nxa, Bacchetta:2022awv} also found difficulty in obtaining agreement, which may indicate an experimental data issue.
Moreover, in contrast to Refs.~\cite{Vladimirov:2019bfa, Cerutti:2022lmb}, we do not find a normalization issue for the E615 $q_T$-dependent dataset, albeit our kinematic cuts are different.
Note that we use the same normalization parameter for the E615 $q_T$-integrated and the $q_T$-dependent observables.
The mean value for this fitted normalization is $1.02$, which is within the reported $16\%$ uncertainty from the E615 experiment~\cite{Conway:1989fs}.

We find that there is no substantial impact on the collinear pion PDFs from the inclusion of the $q_T$-dependent data in the standard CSS framework that we have adopted.
This indicates that the TMD and collinear regimes are well separated in the data we analyzed, in contrast to the high-energy analysis in Ref.~\cite{Bury:2022czx}, and that the measurements correlate more strongly with TMDs than collinear PDFs.
Recent studies~\cite{Gonzalez-Hernandez:2022ifv,Ebert:2022cku} have proposed improvements of the TMD framework, and corresponding implementations and phenomenological analyses will be left to a future work.

{\it Results and discussion.}---\
By definition, the TMD PDF is a 2-dimensional number density dependent on $x$ and~$b_T$.
From Bayes' theorem we can define a conditional density $\tilde{f}_{q/\mathcal{N}}(b_T|x)$ dependent on ``$b_T$ given $x$'' in terms of the ratio
\begin{align}
\tilde{f}_{q/\mathcal{N}}(b_T|x;Q,Q^2) \equiv \frac{\tilde{f}_{q/\mathcal{N}}(x,b_T;Q,Q^2)}{\int \diff^2 {\bm b}_T\, \tilde{f}_{q/\mathcal{N}}(x,b_T;Q,Q^2)}\,.
\end{align}
Notice that this conditional probability is normalized such that $\int \diff^2 {\bm b}_T\tilde{f}_{q/\mathcal{N}}(b_T|x;Q,Q^2) =1$.

\begin{figure}[htb]
    \centering
    \includegraphics[width=0.45\textwidth]{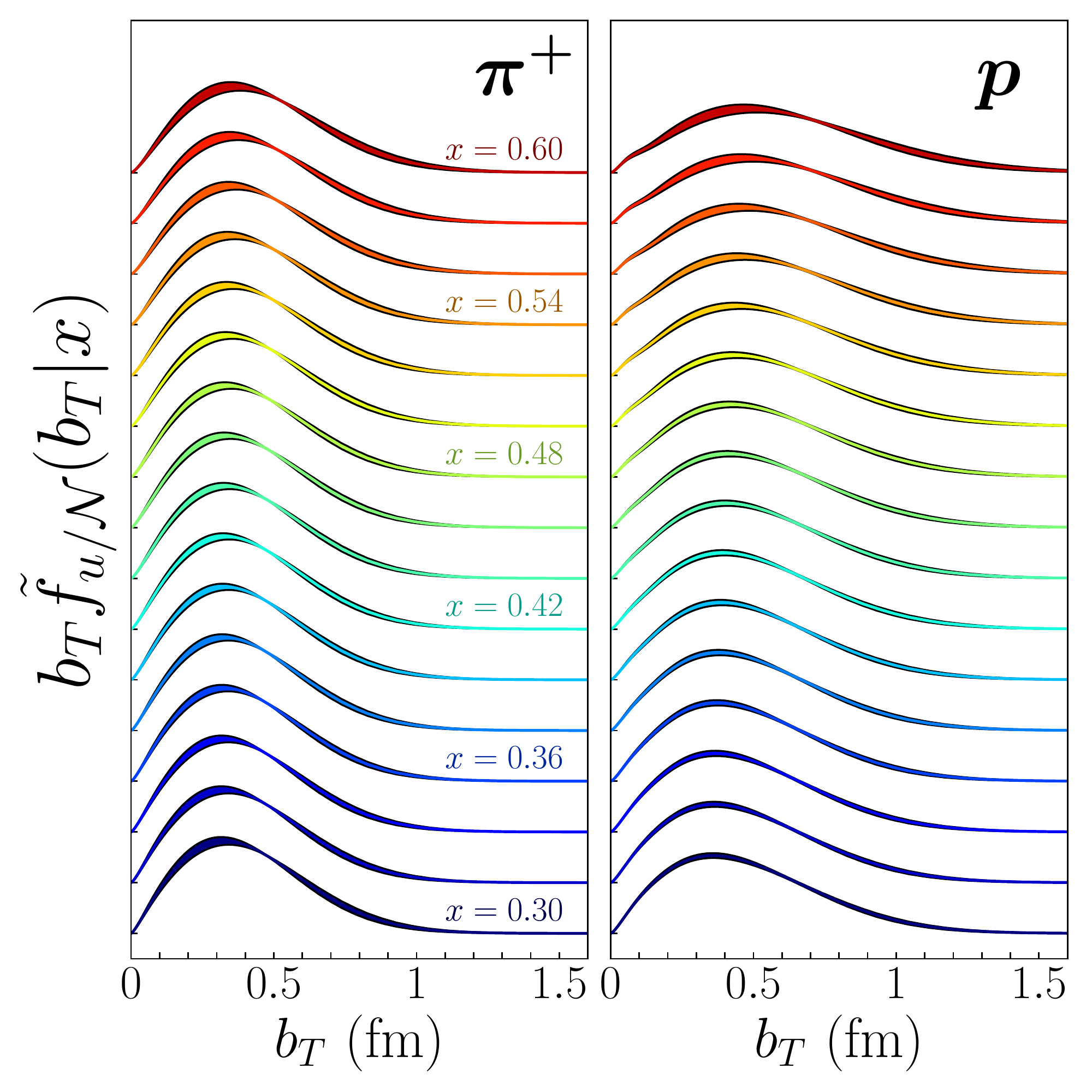}
    \vspace*{-0.3cm}
    \caption{The conditional TMD PDFs for the pion ({\bf left}) and proton ({\bf right}) as a function of $b_T$ for various $x$ values (indicated by color) evaluated at a characteristic experimental scale $Q=6~{\rm GeV}$.
    Each of the TMD PDFs are offset for visual purposes.}
    \label{f.TMDPDF}
\end{figure}

\begin{figure}
    \centering
    \includegraphics[width=0.45\textwidth]{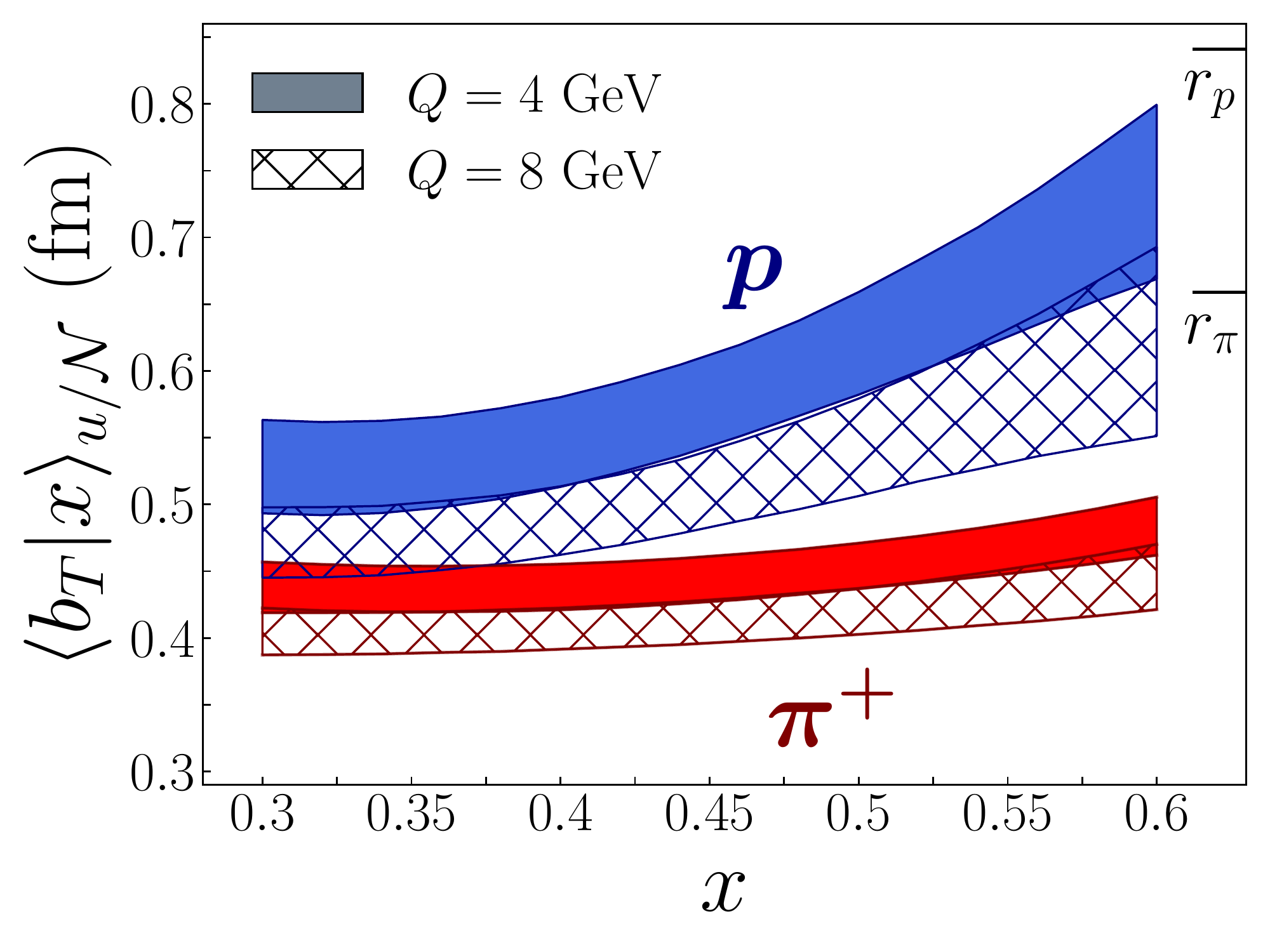}
    \vspace*{-0.3cm}
    \caption{The conditional average $b_T$ calculated from \eref{bTavg} for the $u$ quark in the proton (upper, blue) and in the pion (lower, red) for two $Q$ values as a function of $x$.  The charge radii $r_p$ and $r_\pi$ for each hadron are included for reference~\cite{ParticleDataGroup:2022pth}.
    }
    \label{f.bTavg}
\end{figure}

We show in~\fref{TMDPDF} the extracted  proton and pion conditional densities for the $u$-quark $\tilde{f}_{u/\mathcal{N}}(b_T|x;Q,Q^2)$ in the region covered by the experimental data $x\in[0.3,0.6]$.
Each TMD PDF is shown with its 1$\sigma$ uncertainty band from the analysis.
We focus here on the $u$ quark since our analysis does not include flavor separation in the nonperturbative contribution to the TMDs.
One observes that the pion TMD PDF is significantly narrower in $b_T$ compared with the proton, and both become wider with increasing $x$.
To make quantitative comparisons between the distributions of the two hadrons, we show in Fig.~\ref{f.bTavg} the conditional average $b_T$ as a function of $x$, defined as 
\begin{equation}
\langle b_T | x \rangle_{q/\mathcal{N}} =
\int \diff^2 {\bm b}_T\, b_T\, \tilde{f}_{q/\mathcal{N}}(b_T|x;Q,Q^2),
    \label{e.bTavg}
\end{equation}
for the $u$ quark.
On average there is $\approx 20\%$ reduction of the $u$-quark transverse correlations in pions relative to protons within a $\sim (4-5.2)\,\sigma$ confidence level.
Interestingly,  the charge radius of the pion is also about $20\%$ smaller than that of the proton, using the nominal PDG values ($r_p = 0.8409(4)~{\rm fm}$, 
 $r_\pi = 0.659(4)~{\rm fm}$)~\cite{ParticleDataGroup:2022pth}.
Also, within each hadron, the average spatial separation of quark fields in the transverse direction does not exceed its charge radius, as shown on the right edge of \fref{bTavg}.
Similar qualitative comparative results are shown for $k_T$-space in Refs.~\cite{Bacchetta:2022awv,Cerutti:2022lmb}.

As $x\to 1$, the phase space for the transverse motion $k_T$ of partons becomes smaller, since most of the momentum is along the light-cone direction, and one expects an increase in the transverse correlations in $b_T$ space.
Furthermore, as $Q$ increases more gluons are radiated, which makes TMD PDFs wider in $k_T$ space and therefore narrower in $b_T$ space.
Both of these features are quantitatively confirmed by our results in~\fref{bTavg}.
Importantly, we have checked that the differences between the proton and pion $\langle b_T|x \rangle$ are completely due to the nonperturbative TMD structure, independent of the collinear PDFs, by varying the collinear pion and proton PDF sets to xFitter~\cite{Novikov:2020snp} and MMHT14~\cite{Thorne:2014osa}, respectively, and seeing no difference in Figs.~\ref{f.TMDPDF},~\ref{f.bTavg}.

In Ref.~\cite{Schweitzer:2012hh} it was proposed that $q\bar{q}$ pairs can emerge nonperturbatively through the dynamical breaking of chiral symmetry, which limits the range of the transverse correlations of the quark fields in the hadron.
We can heuristically describe the following physical interpretation of the average $\langle b_T | x \rangle_{u/\cal N} $. In the valence quark dominated regime at large-$x$, valence quarks are occupying the space corresponding to roughly a disc of radius $r_p$. 
The condensate of quark-antiquark pairs in the vacuum arises from gauge field configurations of characteristic size much 
smaller than $r_p$~\cite{Schweitzer:2012hh}. This would imply that once sea quarks emerge in the wave function, $\langle b_T |x \rangle_{u/\cal N}$ should decrease with decreasing values of $x$, which is evident from our findings in~\fref{bTavg}.
However, more work is needed in order to better understand the connection between dynamical chiral symmetry breaking and TMDs in QCD.

\begin{figure}
    \centering
    \includegraphics[width=0.45\textwidth]{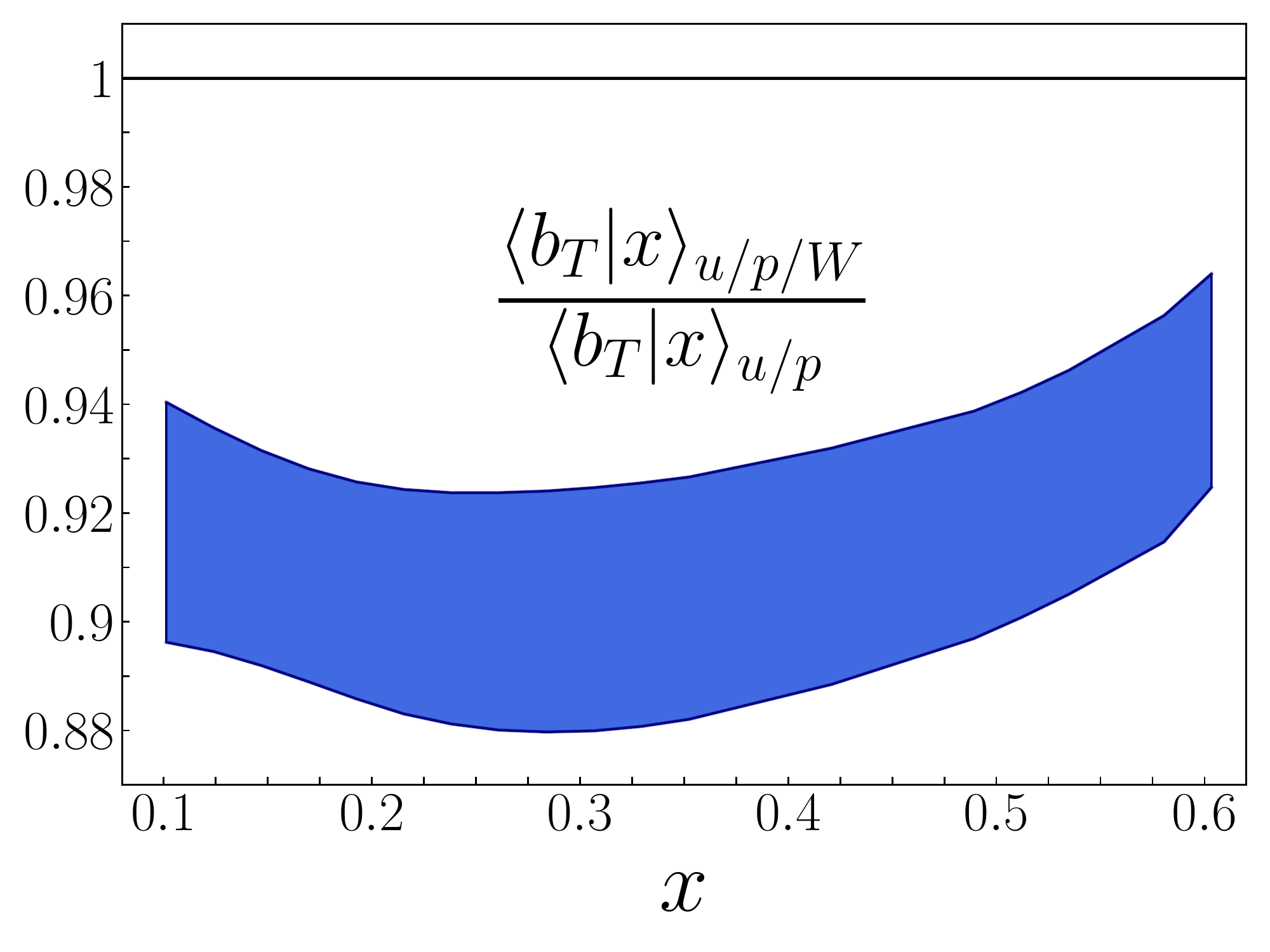}
    \vspace*{-0.3cm}
    \caption{The ratio of the conditional average $b_T$ of the $u$ quark in a proton bound in a tungsten nucleus to that of the free proton at $Q=4~{\rm GeV}$.
    }
    \label{f.nucratio}
\end{figure}

In Fig.~\ref{f.nucratio} we analyze the effect of the nuclear environment on the transverse correlations of quarks inside nucleons, {\it i.e.}, a possible transverse EMC effect, by taking the ratio of $\langle b_T|x \rangle$ for a bound proton in a nucleus  to that of a free proton.
We find an analogous suppression at $x\sim 0.3$, similar to that found in the collinear distributions~\cite{EuropeanMuon:1983wih}.
We have verified that this effect is genuinely produced by the nonperturbative nuclear dependence in the TMD and not from the collinear dependence in the OPE by substituting nCTEQ15~\cite{Kovarik:2015cma} for the EPPS16 nuclear PDFs, and seeing no difference in Fig.~\ref{f.nucratio}.
Additionally, if $a_N$ is set to 0, this ratio is consistent with 1.
Our results are consistent with the earlier findings of Alrashed {\it et al.}~in Ref.~\cite{Alrashed:2022jlx}, but we have gone beyond their study by considering the $x$ dependence of the nonperturbative transverse structure within a simultaneous collinear and TMD QCD global analysis framework.

{\it Conclusions.}---\
We have presented a comprehensive analysis of proton and pion TMD PDFs at N$^2$LL perturbative precision using fixed-target DY data. 
This analysis for the first time used both $q_T$-integrated and $q_T$-differential DY data, as well as LN measurements, to simultaneously extract pion collinear and TMD PDFs and proton TMD PDFs. 
The combined analysis, including an exploration of the nuclear dependence of TMDs, allowed us to perform a detailed comparison of proton and pion TMDs and to study the similarities and differences of their transverse momentum dependence. 

We have determined conclusively that the transverse correlations of quarks in a pion are $\approx 20\%$ smaller than those in a proton, with a more than $4\sigma$ confidence level.  
The observed characteristic decrease of the average separation of quark fields for decreasing $x$ may indicate the influence of dynamical chiral symmetry breaking~\cite{Schweitzer:2012hh}.
This calls for more scrutiny of the connection between our results and these theoretical expectations.
We also found evidence for a transverse EMC effect, as discussed earlier by Alrashed {\it et al.}~\cite{Alrashed:2022jlx}.
We leave for future work the extension of the kinematic region to large $x$, where threshold corrections are needed in both collinear and transverse observables.

The exploration of the quark transverse correlations in pions and protons can be extended to other hadrons, such as kaons and neutrons, in the near future, when the tagged SIDIS programs at Jefferson Lab and the EIC become available.
Such analyses, in combination with future lattice QCD calculations in the TMD sector, will provide a more complete picture of strongly interacting quark-gluon systems that emerge from QCD. \\

{\it Acknowledgments.}---\
We would like to acknowledge many useful discussions with Alexey Vladimirov, Zhongbo Kang, Anatoly Radyushkin, Christian Weiss, and Rabah Abdul Khalek. 
This work has been supported by the U.S. Department of Energy under contracts No. DE-FG02-07ER41460 (LG, EM), No. DE-AC02-06CH11357 (EM, PB), No.~DE-AC05-06OR23177 (PB, WM, AP) under  which Jefferson Science Associates, LLC, manages and operates Jefferson Lab, the National Science Foundation under Grants No.~PHY-2011763 and No.~PHY-2308567 (DP) and No.~PHY-2012002, No.~PHY-2310031, No.~PHY-2335114 (AP), and within the framework of the TMD Topical Collaboration.
The work of WM was partially supported by the University of Adelaide and the Australian Research Council through the Centre of Excellence for Dark Matter Particle Physics (CE200100008).
The work of NS was supported by the DOE, Office of Science, Office of Nuclear Physics in the Early Career Program.

\bibliography{ref}

\end{document}